\documentstyle[12pt,aaspp4,tighten,flushrt]{article}

\def\.{\mathaccent 95}
\def\beq{\begin{equation}}
\def\eeq{\end{equation}}
\def\leq{\label}
\def\a{\alpha}

\def\ga{\gamma}

\def\ep{\epsilon}

\def\la{\lambda}

\def\Om{\Omega}

\def\frac#1#2{{\textstyle{{#1}\over {#2}}}}

\def\lsim{\mathrel{\rlap{\lower4pt\hbox{\hskip1pt$\sim$}}
    \raise1pt\hbox{$<$}}}
\def\gsim{\mathrel{\rlap{\lower4pt\hbox{\hskip1pt$\sim$}}
    \raise1pt\hbox{$>$}}}
\def\sqr#1#2{{\vcenter{\vbox{\hrule height.#2pt
         \hbox{\vrule width.#2pt height#1pt \kern#1pt
         \vrule width.#2pt}
         \hrule height.#2pt}}}}

\newbox\grsign \setbox\grsign=\hbox{$>$} \newdimen\grdimen \grdimen=\ht\grsign
\newbox\simlessbox \newbox\simgreatbox
\setbox\simgreatbox=\hbox{\raise.5ex\hbox{$>$}\llap
     {\lower.5ex\hbox{$\sim$}}}\ht1=\grdimen\dp1=0pt
\setbox\simlessbox=\hbox{\raise.5ex\hbox{$<$}\llap
     {\lower.5ex\hbox{$\sim$}}}\ht2=\grdimen\dp2=0pt

\def\doublespace {\smallskipamount=6pt plus2pt minus2pt
                  \medskipamount=12pt plus4pt minus4pt
                  \bigskipamount=24pt plus8pt minus8pt
                  \normalbaselineskip=24pt plus0pt minus0pt
                  \normallineskip=2pt
                  \normallineskiplimit=0pt
                  \jot=6pt
                  {\def\smallskip {\vskip\smallskipamount}}
                  {\def\medskip   {\vskip\medskipamount}}
                  {\def\bigskip   {\vskip\bigskipamount}}
                  {\setbox\strutbox=\hbox{\vrule 
                    height17.0pt depth7.0pt width 0pt}}
                  \parskip 12.0pt
                  \normalbaselines}

\font\gkvec=cmmib10                         
\def\bomega{\hbox{{\gkvec\char33}}}                  

\def\lb{\langle}
\def\rb{\rangle}

\def\bw{\bar{\omega}}
\def\bv{\bar V}

\def\ts{\times}
\def\lb{\langle}
\def\rb{\rangle}
\def\curl{\nabla {\ts}}

\def\bbV{\bar {\bf V}}
\def\bfv{{\bf v}}

\def\bfj{{\bf j}}

\def\bfw{{\bomega}}

\def\bfb{{\bf b}}

\def\bbB{\bar{\bf B}}
\def\bbA{\bar{\bf A}}
\def\nb{\nabla}
\def\curl{\nb\ts}

\def\b0{b^{(0)}}
\def\v0{v^{(0)}}
\def\w0{\omega^{(0)}}
\def\bb0{\bfb^{(0)}}
\def\bv0{\bfv^{(0)}}
\def\bw0{\bfw^{(0)}}
\def\bj0{\bfj^{(0)}}

\begin{document}

\centerline{\bf Mean Magnetic Field Generation in Sheared Rotators}
\medskip
\centerline{Eric G. Blackman$^{1,2}$}
\centerline{1. Theoretical Astrophysics, Caltech 130-33 Pasadena CA, 91125}
\centerline{2. Institute for Theoretical Physics, UCSB, Santa Barbara, CA, 
93106}

\centerline {\bf ABSTRACT}

A generalized mean magnetic field induction equation 
for differential rotators is derived, including a compressibility, 
and the anisotropy induced on the turbulent quantities 
from the mean magnetic field itself and a mean velocity shear.
Derivations of the mean field equations
often do not emphasize that there must be anisotropy
and inhomogeneity in the turbulence for mean field growth. 
The anisotropy from shear is the source of a term involving the product of 
the mean velocity gradient and the cross-helicity
correlation of the isotropic parts of the 
fluctuating velocity and magnetic field, $\lb{\bfv}\cdot{\bfb}\rb^{(0)}$.  
The full mean field equations are derived to linear order in mean fields, 
but it is also shown that the cross-helicity term survives to all orders
in the velocity shear.  This cross-helicity term can obviate 
the need for a pre-existing seed mean magnetic field for
mean field growth: though a fluctuating seed field is 
necessary for a non-vanishing cross-helicity,  
the term can produce linear (in time) mean field growth of the toroidal
field from zero mean field. 
After one vertical diffusion time, the cross-helicity term becomes 
sub-dominant and dynamo exponential amplification/sustenance of the mean 
field can subsequently ensue.  
The cross-helicity term should produce odd symmetry in the mean magnetic 
field, in contrast to the usually favored even modes of the dynamo 
amplification in sheared discs.  This may be  
important for the observed mean field geometries of spiral galaxies.
The strength of the mean seed field provided by the cross-
helicity depends linearly on the magnitude of the cross-helicity.

\medskip

{\bf Subject Headings}: magnetic fields; galaxies: magnetic fields;
turbulence; accretion discs.

\vfill
\eject
\centerline {\bf 1. Introduction}

Mean magnetic field dynamo theory has been a leading formalism
to characterize the evolution and origin of large-scale magnetic fields
in stars and Galaxies (e.g. Moffatt 1978; Parker
1979; Krause \& R\"{a}dler 1980; Zeldovich et al. 1983).  The mean field
dynamo 
appeals to a combination of helical turbulence, differential rotation,  and
turbulent diffusion to exponentiate  an initial seed mean magnetic
field.  The standard textbook 
kinematic theory is widely known to be incomplete, 
because it ignores the back  reaction of 
the growing magnetic field on the turbulence driving the field growth.

The extent of dynamo growth quenching from the backreaction is not agreed 
upon (Cowling 1957; Piddington 1981; Zeldovich et al. 1983; 
Vainshtein \& Cattaneo 1992;  Cattaneo \& Hughes 1996; 
Vainshtein 1998; Field et al. 1999).
Field et al. (1999) and Blackman \& Field (1999ab) have suggested that some  
previous criticisms of enhanced suppression might be challenged. 
Blackman \& Field 1999b argue that simulations which employ periodic 
boundary conditions (e.g. Cattaneo \& Hughes 1996) 
cannot test for suppression of dynamo theory because an
upper limit on the required dynamo quantities can be shown
to be strongly restricted when boundary terms are ignored.
While some simulations done with diffusive boundary conditions
do show evidence for a mean-field dynamo in fully non-linear MHD 
(Brandenburg \& Donner 1997), there also exist non-linear analytic models
(Vainshtein 1998) which show more extreme suppression.
The restrictions with respect to actual astrophysical 
systems needs to be investigated and more 
analytic and numerical studies are pending.

Since the backreaction consequences are not fully resolved, 
and because some results of backreaction studies do not 
support catastrophic quenching of the dynamo theory, there remains
ample motivation to continue studying the solutions of the 
mean-field dynamo equations in astrophysical systems.
An industry continues to do so, motivated 
by some success in the solar (c.f. Parker 1979; Belvedere 1990), 
galactic (c.f. Ruzmaikin et al 1988; Beck et al 1996), 
and accretion disc (Brandenburg \& Donner 1997)
cases and the simplicity of the theory
(c.f. Moffatt 1978; Parker 1979).
The focus is often on the solutions, rather than the derivation
of the equations being solved.

In this paper, I derive a generalized dynamo equation,
including a restricted compressibility and 
the anisotropy induced from the backreaction 
of the mean magnetic field  and from a mean velocity shear.
The inclusion of the shear induced anisotropy 
leads to an important term in the mean field evolution equations, 
involving the product of mean velocity gradients and
the cross-helicity. This term should 
be included in dynamo models of sheared systems such
as galaxies, stars, and accretion discs, 
as it turns out to play a fundamental role in the mean
field growth as long as the cross-helicity is non-vanishing.
In particular, it can generate a mean magnetic field 
when the initial mean field is zero 
as first pointed out by Blackman \& Chou (1997) and 
Brandenburg \& Urpin (1998).

Previous studies have recognized the potential importance of
the cross-helicity term. Yoshizawa and Yokoi (1993) and Yokoi (1996)
first discussed the term in the astrophysical
context.  Brandenburg \& Urpin (1998) have produced 
a nice analysis of the dynamo equation with shear.
The results herein are closer to those of 
Brandenburg \& Urpin (1998).  However, there is an ambiguity in their 
derivation of the employed dynamo equations of the type
emphasized by Blackman \& Field (1999a).  It is important
to distinguish the isotropic component of the turbulence from the 
anisotropic component, both of which must be present.
The formalism herein explicitly avoids this ambiguity and provides a more 
complete derivation of the generalized dynamo equations with the
important shear term.  In addition, it will be shown 
that the cross-helicity term survives to all orders in the shear not 
just to linear order. 
 
In section 2 the basic mean field
equations are derived including anisotropies induced
both from the mean magnetic field and the velocity shear and compressibility.
In section 3, I show that the term which will lead to linear
mean field growth from zero mean field survives all orders
in the shear anisotropy.
In section 4 the mean magnetic field equation is solved 
in the early time limit when the shear term dominates.
After one vertical diffusion time from the chosen $t=0$, this
term becomes sub-dominant. The resulting solutions are then
given for both $\a-\Om$ and $\a^2$ dynamos.  
Section 5 discusses the implications
of these solutions, with emphasis on galaxies, and
section 6 is the conclusion.


\centerline{\bf 2. Mean field evolution equation}

Mean field dynamo theory 
characterizes the growth of magnetic fields on scales 
larger than the input turbulence.  
The induction equation describing the evolution of the magnetic
field is given by
\beq
\partial_t{\bf B}=\curl({\bf V}\ts{\bf B}) +\lambda \nabla^2{\bf B},
\label{b1}
\eeq
where $\la$ is the micro-physical diffusivity 
and the pressure gradient and is taken parallel to 
the density gradient to ignore 
the Biermann-Battery type term (Biermann 1950).
The Navier-Stokes equation describing the velocity evolution is given by 
\beq
\rho\partial_t{\bf V}=-{\bf V}\cdot\nabla {\bf V}-\nabla P -\nabla (B^2/8\pi)
+{\bf B}\cdot \nabla {\bf B}/(4\pi) + {\bf F}(x,t) + \nabla \phi,
\leq{b2}
\eeq
where $\bf V$ is the velocity, $\bf B$ is the magnetic field, 
$P$ is the pressure, and $\bf F$ is a
forcing function, and $\nabla{\phi}$ is a gravitation term.
Splitting the magnetic field into mean and fluctuating components we 
write $\bbB=\bfb+\bbB$ where $\lb\bfb\rb=0$, and the velocity ${\bf V}$
is similarly defined.
(Note that since astrophysical rotators usually have turbulent
input scales that are a non-trivial fraction of the mean
scales, it is important to distinguish between
fluctuating quantities and small scale quantities.
Their magnitudes  differ by a quantity that varies as
the 3/2 power of the ratio of the turbulent input scale to the mean scale.)

We assume that the mean velocity is time independent,
so we do not consider the mean velocity equation any further.
For the mean induction equation we have (e.g. Moffatt 1978)
\beq
\partial_t \bbB = \curl \lb\bfv\ts\bfb\rb + \curl(\bbV\ts \bbB), 
\leq{b3}
\eeq
where the brackets and the over-bar indicate mean values
and the lower case indicates fluctuating values. 

To write the fluctuating components, I follow the approach 
of Field et al. (1999) and Blackman \& Chou (1997). 
The fluctuating fields are written  
in terms of the zeroth order isotropic
base state and higher order anisotropic components which are due
to the presence of a mean field.  Here 
we 
consider the influence of both the mean magnetic field
as well as the mean velocity field.  
This approach can be motivated
when the magnitude of the fluctuating anisotropic component
of the velocity and magnetic field are $<$  the
mean velocity and mean magnetic field respectively. 

The zeroth order quantities are not solved for but have the properties
of isotropic homogeneous turbulence.  
The base state is also taken to have a 
reflection asymmetry.  In astrophysical systems this property
can result from the presence of an underlying rotation 
and a density gradient (Moffatt 1978; Krause \& R{\"a}dler 1980).  
One can take the liberty of presuming that the forcing function
acts only on the zeroth order state and feeds it with isotropic but 
reflection asymmetric turbulence.  Since we do not
solve the zeroth order equations, we do not need to specify the
explicit form of the forcing function.
The anisotropy induced from the
mean velocity and mean magnetic field will be considered 
in the higher order components.
Compressibility is allowed for in a restricted sense: the density 
will be taken to have a spatially dependent mean component $\bar \rho$, and a 
fluctuating component which is free of influence of the mean fields.
%

Subtracting (\ref{b3}) from (\ref{b1}) we have for the fluctuating field
\beq 
\partial_t \bfb=\curl(\bfv\ts\bfb)-\curl\lb\bfv\ts\bfb\rb
+\curl(\bfv \ts \bbB)+\curl (\bbV\ts \bfb)
\leq{b4}
\eeq
For the fluctuating Navier-Stokes equation we have
\beq
\begin{array}{r}
{\bar \rho}\partial_t{\bfv}=
-\bfv\cdot\nabla\bbV+
\lb\bfv\cdot\nabla\bfv\rb-\bfv\cdot\nabla\bfv-\nabla p
-\nabla(\bbB\cdot\bfb)/4\pi -\nabla b^2/8\pi+\nabla \lb b^2\rb/8\pi\\
+{\bf F}(x,t)+\bbB\cdot\nabla\bfb/4\pi+\bfb\cdot\nabla\bbB/4\pi
+\bfb\cdot\nabla\bfb/4\pi -\lb\bfb\cdot\nabla\bfb\rb/4\pi.
\end{array}
\label{b5}
\eeq
We have ignored the microphsical viscosities for present purposes,
as they are taken to be small.
Write $\bfb=\bfb^{(0)}+\bfb^{(A)}$ 
and similarly $\bfv=\bfv^{(0)}+\bfv^{(A)}$  
where $\bfb^{(A)}$  and $\bfv^{(A)}$  are 
the anisotropic parts of the fluctuating components.
As long as $|\bfb^{(A)}| (\sim 4\pi {\bar \rho}|\bfv^{(A)}|)  < |\bbB|$ and $ |\bbV|$,
then  $\bfv^{(A)}$  and $\bfb^{(A)}$  can in principle,
be solved for explicitly in terms of all orders
in $\bbV$ and $\bbB$ (see Field et al. 1999 for the
solution to all orders in $\bbB$ ignoring mean field gradients).  
In this section I restrict the calculation to linear 
order in both
$|\bbB|/|\bfb|$ and $(|\bbV|/R)/(|\bfb|/L)$, 
which is sufficient for the main points, and then in the next
section I expand to all order in $(|\bbV|/R)/(|\bfb|/L)$,
since this quantity is of order 1 for many applications.

Working in the local standard of rest (LSR) frame (Field et al. 1999)
in which $\bbV=0$, but not $\nabla{\bf {\bar V}}$
to first order in the mean fields we have
\beq 
\begin{array}{r}
\partial_t \bfb^{(1)}=\bfb^{(0)}\cdot\nabla \bfv^{(1)}+
\bfb^{(1)}\cdot\nabla\bfv^{(0)}-\bfb^{(0)}\nabla \cdot\bfv^{(1)}
-\bfb^{(1)}\nabla\cdot\bfv^{(0)}-\curl\lb\bfv\ts\bfb\rb^{(1)} \\
-\bfv^{(0)}\cdot\nabla\bfb^{(1)}-\bfv^{(1)}\cdot\nabla\bfb^{(0)} 
+\bfv^{(0)}\nabla\cdot\bfb^{(1)}+\bfv^{(1)}\nabla\cdot\bfb^{(0)} \\
+\bbB\cdot\nabla\bfv^{(0)}-\bbB\nabla\cdot\bfv^{(0)}-\bfv^{(0)}\cdot\nabla
\bbB
+\bfb^{(0)}\cdot\nabla\bbV-\bfb^{(0)}\nabla\cdot\bbV, 
\end{array}
\leq{b4a}
\eeq
and
\beq
\begin{array}{r}
{\bar \rho}\partial_t{\bfv}^{(1)}=
-\bfv^{(1)}\cdot\nabla\bbV+\lb\bfv\cdot\nabla\bfv\rb^{(1)}-\bfv^{(0)}
\cdot\nabla\bfv^{(1)}-\bfv^{(1)}\cdot\nabla\bfv^{(0)} \\
-\nabla p^{(1)}-\bfv^{(0)}\cdot\nabla\bfv^{(1)} 
-{\bar B}_i \nabla \bfb_i^{(0)}/4\pi  
-b^{(0)}_m\nabla b^{(1)}_m/4\pi-b^{(1)}_m\nabla b^{(0)}_m/4\pi \\
-b^{(0)}_m\nabla {\bar B}_m/4\pi 
+\nabla\lb b^2\rb^{(1)}/4\pi+\bbB\cdot\nabla\bfb^{(0)}/4\pi
+\bfb^{(0)}\cdot\nabla\bbB/4\pi+ \\
(\bfb\cdot\nabla\bfb)^{(1)}/4\pi
-\lb\bfb\cdot\nabla\bfb\rb^{(1)}
/4\pi.
\end{array}
\label{b5a}
\eeq
Now
\beq
\lb\bfv\ts\bfb\rb^{(1)}=\lb\bfv^{(0)}\ts\int\partial_t\bfb^{(1)}dt\rb
-\lb\bfb^{(0)}\ts\int\partial_t\bfv^{(1)}dt\rb. 
\label{b6}
\eeq
Using (\ref{b4a}) and (\ref{b5a}) in (\ref{b6}) we have
for the first term on the right of (\ref{b6})
\beq
\begin{array}{r}
\lb\bfv^{(0)}\ts\int\partial_t\bfb^{(1)}dt\rb_i=
\tau_c\lb\ep_{ijk}v_j^{(0)}\partial_mv_k^{(0)}\rb{\bar B}_m
-\tau_c\lb\ep_{ijk}v_j^{(0)}\partial_m v_m^{(0)}\rb{\bar B}_k\\
-\tau_c\lb\ep_{ijk}v_j^{(0)}v_m^{(0)}\rb\partial_m{\bar B}_k
-\tau_c\lb\ep_{ijk}v_j^{(0)}\partial_m^{(0)}b_k^{(0)}\rb\partial_m{\bar V}_m \\
+\tau_c\lb\ep_{ijk}v_j^{(0)}b_m^{(0)}\rb\partial_m{\bar V}_k
-\tau_c\lb\ep_{ijk}v_j^{(0)}b_k^{(0)}\rb\partial_m{\bar V}_m.
\end{array}
\label{b7}
\eeq
After isotropizing (isotropizing means
that 2 and 3-index tensor average correlations of zeroth order quantities
are proportional to $\delta_{ij}$ and $\epsilon_{ijk}$ tensors respectively) 
and making the controversial (e.g. Cattaneo 1994) but commonly employed 
replacement  the time integrals with
multiples of the correlation time, we have
\beq 
\lb\bfv^{(0)}\ts\int\partial_t\bfb^{(1)}dt\rb=
{\tau_c\over 3}(\lb\bfv^{(0)}\cdot\bfb^{(0)}\rb\curl\bbV
-\lb\bfv^{(0)}\cdot\bfv^{(0)}\rb\curl\bbB
-\lb\bfv^{(0)}\cdot\curl\bfv^{(0)}\rb\bbB)
\label{bfa2}
\eeq
Following the same procedure as above, 
I obtain for the second term on the right of (\ref{b6}) 
\beq 
\begin{array}{r}
-\lb\bfb^{(0)}\ts\int\partial_t\bfv^{(1)}dt\rb=
{\tau_c \over 3}(
\lb\bfb^{(0)}\cdot\bfv^{(0)}\rb\curl\bbV 
+2\lb\bfb^{(0)}\cdot\curl\bfb^{(0)}\rb \bbB/( 4\pi{\bar \rho})\\
-2\lb\bfb^{(0)}\cdot\bfb^{(0)}\rb\curl\bbB/( 4\pi{\bar \rho})
+{\bar \rho}^{-1}\lb \bfb \ts \nabla p^{(1)}\rb).
\end{array}
\label{bf7a}
\eeq
Expand  the last term in (\ref{bf7a}) to first order:  
the energy equation can be written 
\beq
\partial_t P+{\bf V}\cdot\nabla P=
\gamma P\nabla\cdot{\bf v}+{\bar C}
\label{bf7b},
\eeq
where $\ga$ is the adiabatic index and $\bar C$ is a cooling
or heating term which we assume to have only a mean contribution.
Writing the fluctuating part 
of this equation to first order in the mean fields then gives
\beq
\partial_t p^{(1)}=
-\bfv^{(0)}\cdot\nabla{\bar P}+
\lb\bfv\cdot\nabla p\rb^{(1)}
-(\bfv^{(0)}\cdot\nabla p)^{(1)}
+\ga{\bar P}\nabla\cdot\bfv^{(0)}
+\gamma p\nabla\cdot\bbV
+\ga (p\nabla\cdot\bfv)^{(1)}-\ga\lb p\nabla\cdot\bfv\rb^{(1)}.
\label{bf7c}
\eeq
Plugging this back into the last term of (\ref{bf7a}) and using
$\nabla\cdot {\bar {\bf V}}=0$ gives 
\beq
\begin{array}{r}
{\bar \rho}^{-1}\lb \bfb \ts \nabla p^{(1)}\rb_i
=(\lb\ep_{ijk}b_j^{(0)}\partial_k\int\partial_t p^{(1)} dt'\rb
\simeq\tau_c(
(-\ep_{ijk}\lb b_j^{(0)}\partial_k \partial_m p^{(0)}\rb {\bar V}_m
-\ep_{ijk}\lb b_j^{(0)}\partial_k v_m^{(0)}\rb \partial_m {\bar P }\\
-\ep_{ijk}\lb b_j^{(0)} v^{(0)}_m\rb \partial_k \partial_m {\bar P}\\
+\ga\ep_{ijk}\lb b_j^{(0)} \partial_k\partial_m v^{(0)}_m\rb  {\bar P}
+\ga\ep_{ijk}\lb b_j^{(0)} \partial_m v^{(0)}_m\rb
\partial_k{\bar P}).
\end{array}
\label{b7d}
\eeq
The 1st 3rd and 4th terms on the right 
of (\ref{b7d}) vanish from anti-symmetrization after isotropization.
The 5th term vanishes
directly from isotropy of zeroth order correlations.
Thus the only remaining term is the 2nd term.
Combining this term with (\ref{bfa2}) and
(\ref{bf7a}) gives
\beq
\begin{array}{r}
\lb\bfv\ts\bfb\rb^{(1)}=
{\tau_c^2 \over 3{\bar \rho}}\lb\bfb\cdot\curl\bfv\rb^{(0)}\nabla{\bar P}
+(\tau_c\bbB/3)(2\lb\bfb\cdot\curl\bfb\rb^{(0)}/(4\pi{\bar \rho})\\
-\lb\bfv\cdot\curl\bfv\rb^{(0)})
-\tau_c\curl\bbB(\lb\bfb\cdot\bfb\rb^{(0)}/(6\pi{\bar \rho})+{\tau_c \over 3}\lb\bfv\cdot\bfv\rb^{(0)})+{2\tau_c\over 3}\curl\bbV(\lb\bfv\cdot\bfb\rb^{(0)})\\
\equiv\a_1^{(0)}\nabla P+\a_2^{(0)}\bbB-\beta_1^{(0)}\curl\bbB
+\beta_2^{(0)}\curl\bbV\\
\end{array}
\label{b8}
\eeq
Taking the curl of (\ref{b8}), plugging into (\ref{b3}), and
again assuming $\nabla{\bar \rho}\ts\nabla{\bar P}=0$
the equation for the evolution of the mean magnetic field becomes
\beq
\begin{array}{r}
\partial_t\bbB=
\tau_c(\lb\bfb\cdot\curl\bfb\rb^{(0)}/(6\pi{\bar \rho})-\lb\bfv\cdot
\curl\bfv\rb^{(0)}/3)\curl\bbB\\
+\tau_c(\lb\bfb\cdot\bfb\rb^{(0)}/(6\pi{\bar \rho})-\lb\bfv\cdot\curl\bfv\rb^{(0)}/3)\nabla^2\bbB-\tau_c/(6\pi{\bar \rho}^2)\lb\bfb\cdot\curl\bfb\rb^{(0)}
\nabla{\bar \rho}\ts\bbB \\
+\tau_c\lb\bfb\cdot\bfb\rb^{(0)}\nabla{\bar \rho}\ts(\curl \bbB) 
+ {{2\tau_c}\over 3}\lb\bfb\cdot\bfv\rb^{(0)} \curl \curl \bbV+\curl(\bbV\ts\bbB)\\
=
\a_2^{(0)}\curl\bbB+\nabla\a_2^{(0)}\ts\bbB
+\beta_1^{(0)}\nabla^2\bbB-\nabla\beta_1^{(0)}\ts(\curl \bbB)-\beta_2^{(0)}\nabla^2\bbV+\bbB\cdot\nabla\bbV
\end{array}
\label{b10}
\eeq
The last term of (\ref{b10}) 
is the standard $\Omega$ shearing term and
the $\a_2^{(0)}$ and $\beta_1^{(0)}$ terms
are the standard $\a$ and $\beta$
dynamo coefficients (Moffatt 1978)
except for the inclusion of a spatially dependent density.
We will see that the penultimate term on the right 
can generate a seed field. 
The coupling coefficient $\beta_2^{(0)}$ is the cross-helicity.  
The importance of a cross-helicity term was first
addressed by Yoshizawa and Yokoi (1993), Yokoi (1996)
but not as a source of seed field for the standard dynamo.  
Blackman \& Chou (1997) and Brandenburg \& Urpin (1998)
first pointed out the seed field role.  Before solving  
the generalized mean-field equation which
includes the role of the cross-helicity, I will show that the cross-helicity term survives to all orders in the shear.

\centerline{\bf 3. Survival of the linear growth term to all orders in mean shear}

The expansion in section 2, was taken to linear order in 
$|\bbB|/|\bfb^{(0)}|$ and $(|\bbV|/R)/(|\bfb^{(0)}|/L)$, where
$R$ is the scale of variation of $\bbV$ and $L$ is the turbulent
outer scale. 
While $|\bbB|/|\bfb^{(0)}|<1$, $<<1$ for early dynamo evolution, 
and $<1$ in the Galaxy at present, and arguably $<1$ in accretion
discs, the quantity $(|\bbV|/R)/(|\bfb^{(0)}|/L)\simeq 1$ 
in accretion discs for when a shearing instability
drives the turbulence (Balbus \& Hawley 1998).
This is simply the statement that the instability growth time is
of order the rotation time. For this purpose it is necessary to
expand to all orders in at least $(|\bbV|/R)/(|\bfb^{(0)}|/L)$.

In this section I show that the second last term found in (\ref{b10}) 
survives to all orders in the mean shear field
when the $\bbV$ is dominated by 1 component (${\bar V}_\phi$ in the present
case).
Splitting up the small scale fields into isotropic and anisotropic
components we recall
$\bfb=\bfb^{(0)}+\bfb^{(A)}$ and $\bfv=\bfv^{(0)}+\bfv^{(A)}$.  
We then have for the turbulent EMF
\beq
\lb\bfv\ts\bfb\rb=\lb\bfv^{(0)}\ts\bfb^{(A)}\rb+
\lb\bfv^{(A)}\ts\bfb^{(0)}\rb+\lb\bfv^{(A)}\ts\bfb^{(A)}\rb.
\leq{n4.1}
\eeq
To simply show that the linear growth term survives to non-linear
order in the shear, let us take the 
mean magnetic field to be identically zero and
ignore the pressure term. (This will have to be generalized in the
future, but the issue of survival of the term in question is
not affected.)
Then, the evolution equations for the nth order anisotropic
components become (ignoring second order correlations in $\tau_c$) simply 
\beq
\partial_t{v_i}^{(n)}=-v_j^{(n-1)}\partial_j{\bar V}_i,
\leq{n4.2}
\eeq
and
\beq
\partial_t b_i^{(n)}=b_j^{(n-1)}\partial_j{\bar V}_i.
\leq{n4.3}
\eeq
Summing over $n$ we have
\beq
\partial_t v_i^{(A)}+v^{(A)}_j\partial_j{\bar V}_i=
-v_j^{(0)}\partial_j{\bar V}_i,
\leq{n4.4}
\eeq
and
\beq
\partial_t b_i^{(A)}-b^{(A)}_j\partial_j{\bar V}_i=
b_j^{(0)}\partial_j{\bar V}_i.
\leq{n4.5}
\eeq
Multiplying by an integrating factor, 
and defining $M_{iq}(t)\equiv [{\rm e}^{\nabla{\bar {\bf V}}t}]_{iq}$
(where $\nabla {\bar {\bf V}}$ is independent of time)
these equations can be written
\beq
\partial_t(v_i^{(A)}M_{iq}(t))
=-v_j^{(0)}\partial_j{\bar V}_i M(t)_{iq},
\leq{n4.6}
\eeq
and
\beq
\partial_t(b_i^{(A)}M_{iq}(-t))
=b_j^{(0)}\partial_j{\bar V}_i M_{iq}(-t).
\leq{n4.7}
\eeq 
The solutions to these equations are given by  
\beq
v_s^{(A)}=v_i^{(A)}M_{iq}(t)M^{-1}_{qs}(t)
=-\int_0^t v_j^{(0)}\partial_j{\bar V}_i
M_{iq}(t')dt' M_{qs}^{-1}(t)+ v_s(0),
\leq{n4.8}
\eeq
and 
\beq
b_s^{(A)}=b_i^{(A)}M_{iq}(-t)M_{qs}^{-1}(-t)
=\int_0^t b_j^{(0)}\partial_j{\bar V}_i
M_{iq}(-t')dt'M_{qs}^{-1}(-t) + b_s(0)
\leq{n4.9}
\eeq
The 3 terms in the turbulent EMF (\ref{n4.1}) then become
\beq
\lb\bfv^{(0)}\ts\bfb^{(A)}\rb_p=
\lb\ep_{pms}v_m^{(0)}\int b_j^{(0)}
M_{iq}(-t')dt'\rb\partial_j{\bar V}_i
M_{qs}^{-1}(-t)=\lb\ep_{pmi}v_m^{(0)}b_j^{(0)}\rb\partial_j{\bar V}_i
={\tau_c\over 3}\lb\bfv^{(0)}\cdot\bfb^{(0)}\rb(\curl{\bbV})_p
\leq{n5.0}
\eeq
and
\beq
-\lb\bfb^{(0)}\ts\bfv^{(A)}\rb_p=
\lb\ep_{pms}b_m^{(0)}\int v_j^{(0)}
M_{iq}(t')dt'\rb\partial_j{\bar V}_i
M^{-1}_{qs}(t)
=\lb\ep_{pmi}v_m^{(0)}b_j^{(0)}\rb\partial_j{\bar V}_i
={\tau_c\over 3}
\lb\bfv^{(0)}\cdot\bfb^{(0)}\rb(\curl\bbV)_p
\leq{n5.1}
\eeq
and finally
\beq
\begin{array}{r}
\lb\bfv^{(A)}\ts\bfb^{(A)}\rb_p=\lb\ep_{pms}
(-\int_0^tv_j^{(0)}\partial_j{\bar V}_iM_{iq}(t')dt'
M_{qm}^{-1}(t))
(\int_0^t b_l^{(0)}\partial_l{\bar V}_gM_{gw}(-t')dt'
M^{-1}_{ws}(-t)))\rb
+\\
\lb\ep_{pms}v_i^{A}(0)M_{im}(t)
b_j^{(A)}(0)M_{is}(t)\rb.
\end{array}
\leq{n5.2}
\eeq

Notice that the sum of (\ref{n5.0}) and (\ref{n5.1}) 
is simply $\beta_2^{(0)}$ which enters in (\ref{b8})
and thus (\ref{b10}).  Thus (\ref{n5.2}) appears to 
produce an extra term in the EMF.  However, 
it vanishes when $\bbV$ is dominated by 1 component (in our case
${\bar V}_j\simeq {\bar V}_T$, the toroidal mean velocity):  
the index $j$ in $M_{ij}(t)$
always corresponds to the index of ${\bar V}_j$ 
when the exponential in $M_{ij}$ is expanded.  This means that $m=s$ 
in (\ref{n5.2}) and due to the $\ep_{mps}$,
both terms in (\ref{n5.2}) vanish.
Thus, we see that in the case for which the mean velocity is dominated
by one component, the shearing term contributes 
a term which will lead to linear growth from zero mean field, 
as per the next section, 
even when the anisotropy due to this shear is included to all orders.

\centerline{\bf 3. Growth of the Mean Field}

For simplicity, I will assume that the mean field is axisymmetric.
All of the components of the magnetic field can then be expressed
in terms of the toroidal field ${\bar {\bf B}}_T$ and the
poloidal field $\curl {\bar {\bf A}}_T$,
where $\bar {\bf A}$ is the vector potential.  
The equation for the poloidal mean field 
can then be ``uncurled'' (since an arbitrary gradient term
vanishes for axi-symmetry).  From (\ref{b10}) we then have  
\beq
\partial_t{\bar {\bf A}}_T=\a_2^{(0)}{\bbB}_T-\beta_1^{(0)}\curl\curl{\bar {\bf A}}_T
+\beta_2^{(0)}(\curl\bbV)_T +(\bbV\ts\bbB)_T.
\label{c1}
\eeq
For the toroidal component of $\bbB$ we have 
\beq
\partial_t{\bar {\bf B}}_T=-\a_2^{(0)}\nabla^2{\bar {\bf A}}_T-
(\nabla\a_2^{(0)}\cdot){\bar {\bf A}}_T+\beta_1^{(0)}\nabla^2{\bbB}_T
+(\curl{\bar {\bf A}}_T)\cdot\nabla\bbV+(\nabla\beta_1^{(0)}\cdot
\nabla){\bbB}_T-\beta_2^{(0)}\nabla^2\bbV_T.
\label{c2}
\eeq
Let us now proceed to solve equations $(\ref{c1})$ and $(\ref{c2})$.

\medskip
{\bf 3a. Linear Growth Regime}
\medskip

At early times, the dominant term on the right of
$(\ref{c2})$ is the term ${\bf Q}(r,z)\equiv \beta_2^{(0)}\nabla^2\bbV$,
taken to be independent of time.
Solving $(\ref{c2})$ with only this term gives 
\beq
\bbB_T=-{\bf Q}(r,z)t,
\label{c2a}
\eeq 
if we assume the initial $\bbB$ is negligible.
Plugging this solution into (\ref{c1}),  and solving we obtain, to lowest
order in $t$ 
\beq
\bbA_T\simeq 2H^2\a_2^{(0)}{\bf Q}({r,z})t/\beta_1^{(0)}.
\label{c3}
\eeq 
where $H$ is the characteristic gradient length of the mean magnetic field.
How long does this solution hold approximately?
A typical value of the other terms in (\ref{c1}) and (\ref{c2})
is given by the diffusion terms.
Thus plugging (\ref{c2a}) or (\ref{c3}) into the respective
diffusion terms, with the strongest variation of the magnetic field
taken to be the vertical direction, one finds that the linear growth term 
becomes subdominant at $t\gsim H^2/\beta^{(0)}_1$, 
At this time, the fields have grown to the values 
\beq
\bbB_{T,crit}\simeq -(H^2/\beta_1^{(0)})\beta_2^{(0)}\nabla^2\bbV_T
\label{c4}
\eeq
and 
\beq
\bbA_{T,crit}\simeq H^4/(\beta_1^{(0)})^2
\a_2^{(0)}\beta_2^{(0)}\nabla^2\bbV_T.
\label{c5}
\eeq
Because the fluctuating magnetic field builds up to near 
equipartition with the turbulent kinetic motions with a growth rate 
$\sim$ eddy turnover time, $|\bfb^{(0)}|/(4\pi{\bar \rho})^{1/2}\sim 
|\bfv^{(0)}|$ at times of interest for long term mean field evolution. 
We then have
\beq
{\bar V}_{A,crit}\sim |\bbV_T| (H^2/R^2) \chi^{(0)},
\label{c6}
\eeq
where ${\bar V}_{A,crit}$ is the Alfv\'en speed associated with the
mean magnetic field, 
$\chi^{(0)}=[|\beta_2^{(0)}|(4 \pi {\bar \rho})^{-1/2}]/|\beta_1^{(0)}|$,
and ${\bar V}_{A,crit}$ is the Alfv\'en speed associated with the mean
magnetic field at the time when the linear term becomes
subdominant. 
Note the importance of the ratio of 
cross-helicity magnitude to turbulent diffusivity magnitude.
 
\medskip
{\bf 3b. Dynamo Growth Regime}
\medskip

Now I proceed to solve $(\ref{c1})$ and $(\ref{c2})$ in the 
in the $t>H^2/\beta_1^{(0)}$ limit, when the 
when the third term on the right of 
(\ref{c1}) and the last term on the right of 
(\ref{c2}) can be ignored. 
In this regime, writing (\ref{c1}) and (\ref{c2}) 
in terms of their scalar components in cylindrical coordinates gives 
\beq
\partial_t{\bar A}_\phi=\a_2^{(0)}{\bar B}_\phi+\beta_1^{(0)}\nabla^2{\bar A}_\phi
-\beta_1^{(0)}{\bar A}_\phi/r^2,
\label{c6a}
\eeq
where $\phi$ indicates the toroidal component,
and 
\beq
\partial_t{\bar B}_\phi=-\a_2^{(0)}\nabla^2 
{\bar A}_\phi+\a_2^{(0)}{\bar A}_\phi/r^2-\nabla\a_2^{(0)}
\cdot\nabla {\bar A}_\phi
+\beta_1^{(0)}\nabla^2{\bar B}_\phi-\beta_1^{(0)}{\bar B}_\phi/r^2
-\partial_z {\bar A}_\phi \partial_r{\bar V}_\phi+\nabla\beta_1^{(0)}\cdot\nabla {\bar B}_\phi.
\label{c7}
\eeq 

I assume the case where the vertical 
gradients of $\bbB$ dominate and the radial gradients of
${\bar V}_\phi$ dominate and 
look for solutions of the form
$A_\phi\propto {\rm exp}[nt +i{\bf k}\cdot{\bf x}]$ 
and, $B_\phi=\propto {\rm exp}
[nt +i{\bf k}\cdot{\bf x}]$.
(Given the spatial density gradients in a system, one can be more
accurate, since the  $A_{\phi,crit}$ and $B_{\phi,crit}$
have spatial dependences which depend only on the density gradient. 
For present purposes I take the simpler approach.)
From (\ref{c6a}) and (\ref{c7}), we obtain 
\beq
(n+\beta_1^{(0)}k_z^2){\bar A}_\phi-\a_2^{(0)}{\bar B}_\phi=0,
\label{4.1}
\eeq
and
\beq
(\a_2^{(0)}k_z^2-ik_z\partial_z\a_2^{(0)}
-ik_z\partial_r{\bar V}_\phi){\bar A}_\phi
-(n+\beta_1^{(0)}k_z^2-ik_z\partial_z\beta_1^{(0)}){\bar B}_\phi=0.
\label{4.2}
\eeq
The dispersion relation resulting from 
(\ref{4.1}) and (\ref{4.2})
is
\beq
n^2+n(2\beta_1^{(0)}k_z^2-ik_z\partial\beta_1^{(0)})
+(k_z\beta_1^{(0)})^2-ik_z^3\beta_1^{(0)}\partial_z\beta_1^{(0)}
-(k_z\a_2^{(0)})^2+i\a_2^{(0)}k_z\partial_z\a_2^{(0)}+i\a_2^{(0)}k_z\partial_r
{\bar V}_\phi.
\label{4.3}
\eeq
The solution to this quadratic is
\beq
2n=-k_z \beta^{(0)}_1+ik_z\partial_z\beta^{(0)}_1\pm[4k_z^2{\a_2^{(0)}}^2
-k_z^2(\partial_z\beta_1^{(0)})^2-i4k_z\a_2^{(0)}(\partial_z \a_2^{(0)}+
\partial_r{\bar V}_\phi)]^{1/2}.
\leq{4.4}
\eeq
Scaling the terms inside the square brackets on the right of
(\ref{4.4}) shows that the second term is down from the first by some constant
$\lsim 1$ times $(L/H)^2$, where $L$ is the dominant turbulent
scale.  We assume that this second term can be
ignored for simplicity.
The remaining real part of (\ref{4.4}) is of interest.
This can be found from the following:  write the right
side as $(c+di)^{1/2}$, where $c$ and $d$ are real 
and seek the real quantity $a$ such that 
$a+bi=(c+di)^{1/2}$.  In general, this implies
that 
\beq
2a^2=[c\pm(c^2+d^2)^{1/2}].
\leq{4.5}
\eeq
Applying (\ref{4.5}) to (\ref{4.4}) 
gives
\beq
{\rm Re}[n]=-k_z^2 \beta_1^{(0)}\pm k_z (\a_2^{(0)}/2^{1/2})[
1\pm[1+(\partial_z\a_2^{(0)}
+\partial_r{\bar V}_\phi)^2/(k_z \a_2^{(0)})^2]^{1/2}]^{1/2}.
\leq{4.6}
\eeq

There are two interesting limits of (\ref{4.6}). 
In the limit that the differential rotation dominates
the inner sum, we have 
\beq
{\rm Re}[n]\simeq -k_z^2 \beta_1^{(0)}
\pm (k_z \a_2^{(0)})^{1/2}(\partial_r{\bar V}_\phi)^{1/2}/2^{1/2}.
\leq{4.7}
\eeq
This has growing modes for 
$k_z \lsim  (|\a_2^{(0)}\partial_r{\bar V}_\phi|/2{\beta_1^{(0)}}^2)^{1/3}$.
The maximum growth wavenumber is $k_{z,max}= 
(|\a^{(0)}_2\partial_r{\bar V}_\phi|/2^5{\beta_1^{(0)}}^2)^{1/3}$
so that $n_{max}\sim 0.3 (|\a^{(0)}_2\partial_r{\bar V}_\phi|^2/{\beta_1^{(0)}})^{1/3}$.  This is an ``$\a-\Om$'' type dynamo (c.f. Moffatt 1978; Ruzmaikin
et al. 1988).

In the limit that $\partial_z\a_2^{(0)}$ dominates or is comparable
to $\partial_r{\bar V}_\phi$, we have
\beq
{\rm Re}[n]\simeq -k_z^2 \beta_1^{(0)}\pm k_z \a_2^{(0)}
[1\pm [1+(\partial_z\a_2^{(0)}/k_z \a_2^{(0)})^2]^{1/2}]^{1/2}
\simeq -k_z^2 \beta_1^{(0)}\pm k_z \a_2^{(0)}
[1\pm 2^{-1/2}]^{1/2},
\leq{4.8}
\eeq
where the latter similarity follows when
$|(\partial_z\a_2^{(0)})/(k_z \a_2^{(0)})| \sim 1.$
This has growing modes for $k_z \lsim 1.3 \a_2^{(0)}/\beta_1^{(0)}$.
The maximum growth wave number is at $\sim 0.65 \a_2^{(0)}/\beta_1^{(0)}$
so that $n_{max}\simeq 0.4 ( \a_2^{(0)})^2/\beta_1^{(0)}$.
This is an ``$\a^2$'' type dynamo (c.f. Moffatt 1978).


\centerline{\bf 4. Discussion}

The results of sections 3a and 3b show the two phases of mean field growth
for a shearing (disc) rotator. 
The first regime does not require a seed mean field,
only a fluctuating field whose mean can vanish. 
The toroidal field produced in the first phase (the linear growth
regime) should be asymmetric with respect to the disc plane (like an 
A0 mode, e.g. Beck et al. 1996)
since the cross-helicity $\beta_2^{(0)}=\lb\bfb^{(0)}\cdot\bfv^{(0)}\rb$
is a pseudo-scalar.  The poloidal field growth feeds on this
linear toroidal field.
If the cross-helicity has the same sign throughout a hemisphere, the 
linear growth phase should not
produce radial field reversals in that hemisphere
since the mean field vector is determined
only by the sign of the cross-helicity and the direction of the mean
velocity and its gradient.

The initial growth phase is important since 
standard dynamo models generally consider 
the equations only in the second phase.  In the second phase 
it is generally easier to amplify S0 type modes, whose toroidal field 
is symmetric with respect to the galactic plane (Ruzmaikin et al 1988; 
Beck et al 1996),
though more complicated models are able to excite mixed modes e.g. A0-S0
(Moss \& Tuominen 1990; Moss et al. 1993).
The dominant observed field after amplification saturates also 
depends on which seed field modes 
were present to begin with (Ruzmaikin et al. 1988; Poezd et al. 1993). 
Suppose an asymmetric seed field is present with magnitude
well in excess of the seed symmetric mode.  Then even if the asymmetric
mode grows more slowly than the symmetric mode, the difference
of initial magnitudes can more than compensate and the asymmetric
mode could dominate.  Also, once equipartition is
reached, different modes may oscillate with different frequencies.

In the Galaxy for example, the magnitude of the asymmetric seed resulting 
from the linear growth phase is, assuming no vertical shear and 
a radius of $8$Kpc, from (\ref{c6}) 
${\bar B}\sim 2.7 \ts 10^{-9}
(\chi^{(0)}/0.01)(H/0.5{\rm Kpc})^2(R/8{\rm Kpc})^{-2}(V_\phi/2\ts 10^7{\rm cm/s})$G.
This is a substantial seed field, which would arise in one vertical
diffusion time.  I have scaled with $\chi^{(0)}\sim 0.01$. 
However, note in the Galaxy for example, the dynamo can exponentiate
$>20$ or so times in the  Galactic lifetime.  Thus 
even if $\chi^{(0)} \sim 10^{-5}$ the seed from the linear growth phase 
could be important for seeding the subsequent dynamo growth  
up to its present value of few $\ts 10^{-6}$ G.  
If $\chi^{(0)}\ge 0.01$, or if there were vertical shear, then (\ref{c6})
would imply an even larger seed, and a much shorter time
for the cross-helicity + dynamo to produce an equipartition mean field.
The exponential amplification 
rate of a symmetric mode from a seed field of $10^{-9}$G
would have to be larger than that of  
the asymmetric mode to be competitive with that seeded from 
the asymmetric linear growth phase.

Han et al. (1998) suggest that the Faraday rotation 
data for M31 supports the presence of an even mode (S0).  
For our Milky way however, the data favor a dominant
A0 mode (Han et al. 1997).  While more galaxy data are 
needed, there appears to be the possibility that different
modes dominate in different galaxies.  
The results herein suggest that the cross-helicity term
can be important in determining the field geometry
in the framework of in situ field generation models.

A number of simple but important complications affect the 
observational implications of the above results. 
First, inhomogeneity and large scale local structures
can lead to coherent local structures. 
There may also be seed fields present from
protogalactic, cosmological, or supernovae origins which are 
amplified concurrently.  In addition, the mean field in a galaxy 
is likely never to be zero since the dominant input scale of 
the turbulence in galaxies 
is always  a non-trivial fraction of the galactic radius.
For example, at the solar location of our Galaxy, the scale ratio is at most 
$(100{\rm pc}/10{\rm kpc})$.  Thus, just the random RMS field (Blackman
1998), or RMS contributions to dynamo coefficients that result 
(Vishniac \& Brandenburg 1997) may be important 
in influencing the observation of a mean field.  
(It is important to note that mean field theory 
and Faraday rotation measurements are degenerate
with respect to mean field topology: they cannot distinguish between
a large scale field formed from averaging over small scale loops,
or from that formed by averaging of a topologically connected field
line with fluctuations.  Spectral energy distribution approaches
also share this degeneracy.)  

The two phase growth described herein, as that of Brandenburg \& Urpin (1998), 
could be testable in some non-linear MHD disc simulations.
Note that periodic boundary condition simulations
cannot test for mean field growth because the mean field is conserved
by construction (c.f. Balbus \& Hawley 1998). It is known
that diffusion through the boundary is required for a working
dynamo (Parker 1979; Ruzmaikin et al 1988), 
and it has also been realized that diffusive boundary conditions
are required for a non-vanishing $\alpha_{1}^{(0)}$ 
effect even when dynamo action is 
not present (Blackman \& Field 1999b). 
There is some evidence that a 
dynamo is operating in steady accretion disc simulations with the
appropriate diffusive boundary conditions (Brandenburg \& Donner 1997).

Brandenburg \& Urpin (1998) point out that some  
previous simulations have estimated $\chi$. (Note that 
this is not $\chi^{(0)}$ which comes in directly into the present
formalism, but the full $\chi$, to all orders in the mean fields.
The two should be distinguished.) It is found that $\chi\sim 0.03$
in stratified convection simulations (Brandenburg et. al. 1996), 
$\chi\sim 3\ts 10^{-4}$ in magneto-shearing driven instability simulations
(Brandenburg et. al. 1995), and $\chi \sim 5\ts 10^{-3}$ in supernova induced
turbulent flows (Korpi et al. 1998; though these have not reached the 
saturated steady-state) potentially relevant for galaxies.
Chandran \& Rodriguez (1997) studied the evolution of the cross-helicity 
in the framework of the direct interaction approximation,
and do not rule out the possibility of significant cross-helicity
on the turbulent input scale. 
Future simulations should give a better handle on $\chi$
and $\chi^{(0)}$, and a comparison of the two 
in the various settings. 
As pointed out earlier, even values of $10^{-5}$ might be significant.
Note that it is also important to distinguish between
an RMS value that may fluctuate in time and a systematic value
that remains steady in time.



The use of a linear backreaction model 
is certainly incomplete (as  are most studies of dynamo equation solutions).
Both the shear and $\bbB$ should be included to all orders.
Field et al. (1999) solved for all orders in $\bbB$ without field gradients. 
Ultimately, more analytic and numerical work are needed
to understand how well the dynamo survives in the non-linear backreaction. 
Interestingly, dynamo theory in its standard form is the ``complement''
of analytic accretion on disc theory: the latter ignores the
dynamics of the magnetic field, while the former does not
fully include the magnetic backreaction on the velocity dynamics.
Both require some closure approximation to turbulence to make
analytic progress.  Studies which focus on the solutions
of the accretion disc or mean field dynamo equations often gloss
over the approximations which led to the form of the equations
being solved.


\centerline{\bf 5. Conclusions}

The equations for the evolution of the mean magnetic field 
in a sheared rotator have been derived, taking into account the anisotropy
induced in the turbulence from the large 
scale magnetic field and the differential shear. 
A restricted compressibility has also been included.  
A shear term violating the homogeneity of the
dynamo equations was shown to survive to all orders in the anisotropy.
For finite cross-helicity, the mean magnetic field can grow in two phases.
Linear growth ensues for one vertical diffusion time.
The mean toroidal and poloidal fields 
grow without a seed mean field to a value whose Alfv\'en
speed is of order $\gsim (H/R)^2 \chi^{(0)} V_\phi$ 
in an A0 mode (toroidal field anti-symmetric
with respect to the disc plane). After the vertical diffusion time,
the second phase incurs, and the more ``standard'' dynamo evolution 
takes over.  If the linear growth phase seed field dominated the
field geometry, there would be a reversal in the toroidal field across the disc
plane and no reversals in radius if the cross-helicity has 
the same sign in a given hemisphere.  
Determining the magnitude of $\chi^{(0)}$ and the full $\chi$ to all orders 
for various applications will continue to be aided by numerical simulations. 
For non-vanishing cross-helicity, equation 
(16) should replace the standard textbook dynamo equations for 
sheared rotators.

Acknowledgements: Thanks to the Black Hole Astrophysics Program at the ITP
where part of this work was carried out and supported by NSF Grant 
PHY94-07194, and to the referee, editor, and G. Field for comments.

\end{document}